\documentclass[pdflatex,sn-mathphys-num]{sn-jnl}

\usepackage{graphicx}%
\usepackage{multirow}%
\usepackage{amsmath,amssymb,amsfonts}%
\usepackage{amsthm}%
\usepackage{mathrsfs}%
\usepackage{xcolor}%
\usepackage{textcomp}%
\usepackage{manyfoot}%
\usepackage{booktabs}%
\usepackage{algorithm}%
\usepackage{algorithmicx}%
\usepackage{algpseudocode}%
\usepackage{listings}%
\usepackage[numbers]{natbib}

\theoremstyle{thmstyleone}%
\theoremstyle{thmstyletwo}%

\theoremstyle{thmstylethree}%

\newcommand{\hl}[1]{{ \color{brown}{[HL: #1]}}}

\begin{document}

\title[Article Title]{Lithological Controls on the Permeability of Geologic Faults: Surrogate Modeling and Sensitivity Analysis}

\author*[1]{\fnm{Hannah} \sur{Lu}}\email{hannah.lu@austin.utexas.edu}

\author[2,3]{\fnm{Llu\'{i}s} \sur{Sal\'{o}-Salgado}}\email{lsalo@mit.edu}

\author[3,4]{\fnm{Ruben} \sur{Juanes}}\email{juanes@mit.edu}

\affil*[1]{\orgdiv{Oden Institute for Computational Engineering and Sciences}, \orgname{The University of Texas at Austin}, \orgaddress{\street{201 E 24th St}, \city{Austin}, \postcode{78712}, \state{TX}, \country{USA}}}

\affil[2]{\orgdiv{Department of Earth and Planetary Sciences}, \orgname{Harvard University}, \orgaddress{\street{20 Oxford St}, \city{Cambridge}, \postcode{02138}, \state{MA}, \country{USA}}}

\affil[3]{\orgdiv{Department of Civil and Environmental Engineering}, \orgname{Massachusetts Institute of Technology}, \orgaddress{\street{77 Massachusetts Avenue}, \city{Cambridge}, \postcode{02139}, \state{MA}, \country{USA}}}

\affil[4]{\orgdiv{Department of Earth, Atmospheric, and Planetary Sciences}, \orgname{Massachusetts Institute of Technology}, \orgaddress{\street{77 Massachusetts Avenue}}, \city{Cambridge}, \postcode{02139}, \state{MA}, \country{USA}}

\abstract{Fault zones exhibit complex and heterogeneous permeability structures influenced by stratigraphic, compositional, and structural factors, making them critical yet uncertain components in subsurface flow modeling. In this study, we investigate how lithological controls influence fault permeability using the PREDICT framework: a probabilistic workflow that couples stochastic fault geometry generation, physically constrained material placement, and flow-based upscaling. The flow-based upscaling step, however, is a very computationally expensive component of the workflow and presents a major bottleneck that makes global sensitivity analysis (GSA) intractable, as it requires millions of model evaluations. To overcome this challenge, we develop a neural network surrogate to emulate the flow-based upscaling step. This surrogate model dramatically reduces the computational cost while maintaining high accuracy, thereby making GSA feasible. The surrogate-model-enabled GSA reveals new insights into the effects of lithological controls on fault permeability. In addition to identifying dominant parameters and negligible ones, the analysis uncovers significant nonlinear interactions between parameters that cannot be captured by traditional local sensitivity methods.}

\keywords{Global Sensitivity Analysis, Surrogate Models, Fault Permeability, Stochastic Modeling}



\maketitle
\section{Introduction}
Fault zones in siliciclastic formations exhibit highly heterogeneous and anisotropic permeability structures, which have a strong influence on subsurface fluid flow, pressure propagation, and solute transport~\cite{caine1996,bense2013,vrolijk2016clay,sosioderosa2018,salo2023fault}. Accurate characterization of fault permeability is essential for a wide range of applications, including hydrocarbon recovery, groundwater management, the safe storage of carbon dioxide in geological formations, or geothermal energy extraction~\cite{aydin2000,faulkner2001can,wibberley2003internal,vrolijk2016clay,krevor2023,paldor2024,kivi2024}. However, the spatial variability in fault architecture---driven by lithological control factors such as stratigraphy, burial depth, and dip angle---introduces substantial uncertainty in the permeability field. Understanding how this uncertainty propagates to impact effective fault permeability at the macroscale is critical for understanding of fault formation and managing risk in reservoir-scale operations.

To address these challenges, the recently developed modeling framework, PREDICT (PeRmEability DIstributions of Clay-smeared faulTs)~\cite{salo2023fault}, provides a quantitative description of the lithological controls on fault material distribution and computes probability distributions for the directional components (dip-normal, strike-parallel, and dip-parallel) of the fault permeability tensor from statistical samples for a set of geological variables. The inputs include geometrical, compositional, and mechanical properties that govern the structure of the fault core~\cite{caine1996,bense2013,vrolijk2016clay}. By simulating physically-constrained, stochastic material placement---populating the fault zone with sand and clay smears---PREDICT enables multiple realizations of fault architecture and performs flow-based upscaling to obtain permeability fields that are suitable for reservoir-scale modeling and uncertainty quantification.

Given a prior estimation of the geological input space, a natural next step is to perform sensitivity analysis to quantify how the lithological factors control the fault permeability. Sensitivity analysis provides a quantitative framework for evaluating how input uncertainty propagates to model outputs and for ranking the relative importance of input parameters. Local sensitivity analysis (LSA) explores output responses to small, independent perturbations in each input variable around a nominal baseline. While simple and efficient, LSA assumes linearity and ignores interactions between variables, making it inadequate for capturing the coupled, nonlinear effects often present in geologic systems. In PREDICT, for example, the intermediate geologic variables controlling fault material properties depend nonlinearly on multiple inputs. In contrast, global sensitivity analysis (GSA) evaluates parameter influence across the full input space, capturing nonlinearities and higher-order interactions~\cite{saltelli1993sensitivity,iooss2015review}. Variance-based GSA methods, such as those based on Sobol’ indices, are particularly powerful, as they decompose the total output variance into contributions from individual inputs and their interactions.

However, the application of GSA to the PREDICT framework is severely limited by computational cost. Each model evaluation in PREDICT requires a full realization of the stochastic workflow—including conditional sampling, physically constrained material placement, and, most critically, flow-based upscaling using MATLAB Reservoir Simulation Toolbox (MRST~\cite{lie2019introduction}) on a high-resolution mesh---to obtain $N_\text{sim}$ permeability samples. In practice, $N_\text{sim}$ is typically on the order of one thousand to ensure a meaningful representation of the underlying distribution~\cite{salo2023fault}. Since variance-based GSA typically requires on the order of $(N_p + 2)N_{\text{mc}}$ model evaluations (where $N_p$ is the number of input parameters and $N_{\text{mc}}$ is the number of Monte Carlo samples), the total number of upscaling operations ($(N_p + 2)N_{\text{mc}}N_\text{sim}$) quickly becomes prohibitive. For example, even a modest study with six input parameters, 5,000 Monte Carlo samples and 1,000 permeability samples in each PREDICT model evaluation would require 40 million simulations—each involving a computationally intensive flow-based upscaling runs by MRST. This renders direct GSA with PREDICT intractable and motivates the development of an efficient surrogate model to replace the high-fidelity upscaling step, thereby enabling scalable global sensitivity analysis.

In this work, we address this bottleneck by developing a neural network-based surrogate model to emulate the flow-based upscaling step within the PREDICT framework. We design a convolutional neural network (CNN) based on a UNet architecture~\cite{ronneberger2015u} for image-to-scalar regression, which maps 2D permeability fields to their corresponding upscaled scalar permeability components. The surrogate model is trained on a representative dataset generated using MRST and achieves high accuracy while reducing the cost of individual model evaluations by several orders of magnitude. The resulting surrogate-accelerated workflow becomes a critical enabler for tractable global sensitivity analysis. In contrast to local sensitivity methods that isolate parameters one at a time, our global sensitivity analysis captures nonlinear and interaction effects among parameters, providing a more complete picture of system behavior. This enhanced understanding allows us to identify the dominant lithological controls, quantify secondary influences, and uncover synergistic effects that are otherwise undetectable~\cite{sobol1993sensitivity}.

The remainder of the paper is organized as follows. In Section~\ref{sec: setup}, we briefly review the PREDICT framework and describe the idealized two-layer fault system used in this study. Section~\ref{sec: LSA} presents the results of a local sensitivity analysis to illustrate the individual effects of lithological parameters. In Section~\ref{sec: GSA}, we introduce a neural network-based surrogate model for flow-based upscaling and demonstrate how it enables tractable global sensitivity analysis. We then present the results of the global analysis and compare them with local findings to highlight the importance of nonlinear interactions and coupled effects. Finally, Section~\ref{sec: conclusion} summarizes the key conclusions and discusses future directions for extending this approach to broader geologic settings and making use of the sensitivity analysis results to reservoir scale applications.

\section{Lithological Controls on Fault Permeability}\label{sec: setup}
Faults in siliciclastic sequences exhibit highly variable permeability structures, often with orders-of-magnitude contrasts between the fault core and surrounding damage zones~\cite{caine1996,bense2013,vrolijk2016clay,salo2023fault}. Measurements across a range of host rocks show that fault core permeability in mature settings typically falls between $10^{-17}$ and $10^{-21}$~m$^2$, whereas surrounding fractured zones can reach values of $10^{-14}$ to $10^{-16}$~m$^2$~\cite{faulkner2001can,wibberley2003internal,faulkner2006slip,lockner2009geometry}. In addition, clay smears commonly develop in normally-consolidated, shallow ($< 3$~km depth) fault zones~\cite{vrolijk2016clay}, further complicating the permeability structure. These observations highlight the need to account for strong heterogeneity, anisotropy, and uncertainty in fault permeability modeling, and underscore the importance of understanding how lithological controls influence the resulting fault permeability distributions~\cite{bense2013,vrolijk2016clay,salo2023fault}.

\subsection{Overview of PREDICT}
A new methodology, named PREDICT, was introduced recently in~\cite{salo2023fault} to compute probability distributions for the directional components (dip-normal, dip-parallel, and, in the 3D version, strike-parallel) of the fault permeability tensor from statistical samples for a set of geological variables. These variables, which include geometrical, compositional, and mechanical properties, allow multiple discretizations of the fault core to be populated with sand and clay smears, which can be used to upscale the permeability to a coarser scale. The flexibility of the user-defined upscaling can be used to assign fault permeability in subsequent study of reservoir-scale flow simulations~\cite{salo2025} and uncertainty quantifications~\cite{lu2024uncertainty}.

\begin{figure}[!h]
\centering
    \includegraphics[width = \textwidth]{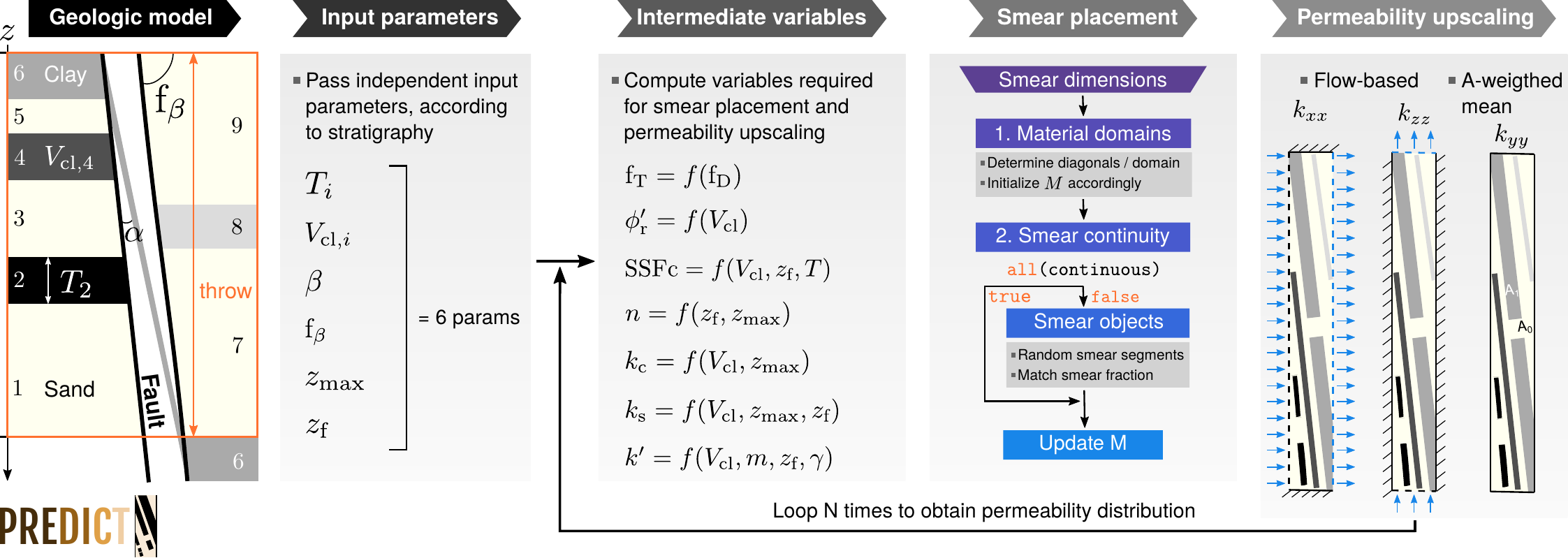}
        \caption{PREDICT's workflow (left to right). Stratigraphic section in the \textcolor{orange}{throw window} of interest is described by input parameters, which PREDICT uses to compute ranges and probability distributions for intermediate variables. For each fault section, fault thickness $f_T$, residual friction angle ($\phi_r^\prime$), critical shale smear factor (SSFc), porosity ($n$), permeability ($k$), and permeability anisotropy ratio ($k^\prime$) are sampled and used to place fault materials and assign their properties. Subscript “c” refers to clay smear, and “s” refers to sand smear. Permeability is upscaled using 2-D fault volume.}
        \label{fig: PREDICT}
\end{figure}

A summary of PREDICT's 2D workflow is shown in Fig.~\ref{fig: PREDICT}. The computation is performed in a given throw window, in which PREDICT represents the fault core. The input parameters describing the faulted stratigraphy include the layer thickness ($T_{i}$), clay volume fraction ($V_{\text{cl},i}$), dip angle ($\beta$), the fault dip ($f_\beta$), maximum burial depth ($z_\text{max}$) and faulting depth ($z_f$). Marginal probability distributions for another set of intermediate geologic variables including fault thickness $f_T$, residual friction angle ($\phi_r^\prime$), critical shale smear factor (SSFc), porosity ($n$), permeability ($k$), and permeability anisotropy ratio ($k^\prime$), are generated and used to place fault materials and assign their properties. Finally, the permeability for the studied throw window is computed in the coarse grid by flow-based upscaling of the fine grid permeability using the MATLAB Reservoir Simulation Toolbox (MRST~\cite{lie2019introduction}). Repeated realizations of this process result in probability distributions for $k_{xx}$ and $k_{zz}$. We refer the readers to \cite{salo2023fault} and its supplemental material for details including geologic range of application, selection of input parameters, and notation.

The goal of this study is to understand how lithological controls—represented by the input parameter vector $\mathbf p \in \mathbb{R}^{N_p}$—influence the resulting fault permeability distributions. Specifically, we aim to quantify how variations in stratigraphic, compositional, and structural properties propagate through the PREDICT framework to affect the statistical distribution of fault permeability in both the dip-normal ($k_{xx}$) and dip-parallel ($k_{zz}$) directions. This relationship is formally expressed as

\begin{equation}\label{eq: model}
Q = \mathcal{M}_\text{PREDICT}(\mathbf{p}),
\end{equation}
where $Q$ denotes the quantities of interest (QoIs), such as selected statistical quantiles of the output permeability distributions, and $\mathcal{M}_\text{PREDICT}$ denotes the forward model implemented by the PREDICT. To evaluate the impact of individual and combined input parameters on the variability of $Q$, we perform both local and global sensitivity analyses (section~\ref{sec: LSA} and~\ref{sec: GSA} respectively), which allow us to identify the dominant lithological controls governing fault permeability uncertainty.

\subsection{A Study Case of a Two-Layer System}
\begin{figure}[!h]
    \includegraphics[trim = 0 700 0 110, clip, width = \textwidth]{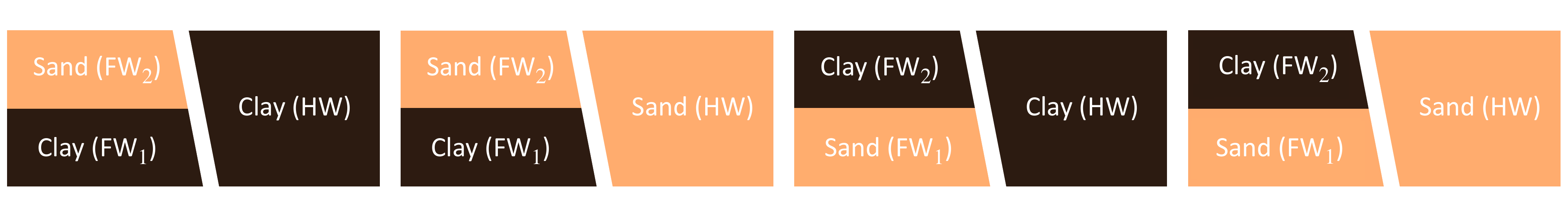}
    \caption{Four scenarios of the two-layer system: CSC, CSS, SCC, and SCS (left to right). The throw window spans 100 meters, with the footwall composed of two 50-meter-thick layers.}
    \label{fig: 2-layer-sys}
\end{figure}

To demonstrate the PREDICT framework and investigate the influence of lithological controls, we consider an idealized two-layer system within a fixed throw window (refer to Fig.~\ref{fig: PREDICT} for throw window definition). In this configuration, the hanging wall consists of a single material layer with a thickness of 100 meters, while the footwall comprises two distinct material layers, each 50 meters thick. Fig.~\ref{fig: 2-layer-sys} illustrates the four possible stratigraphic scenarios for this setup: Clay-Sand-Clay (CSC), Clay-Sand-Sand (CSS), Sand-Clay-Clay (SCC), and Sand-Clay-Sand (SCS). We denote each stratigraphic scenario by listing the layers from the bottom footwall (FW$_1$) upward to the upper footwall (FW$_2$) and then to the hanging wall (HW).

The lithological control parameters considered in this case study include six variables: clay volume fractions in the three layers ($V_\text{cl}^{\text{FW}_1}$, $V_\text{cl}^{\text{FW}_2}$, and $V_\text{cl}^{\text{HW}}$)—following the same order from the bottom footwall to the upper footwall and then to the hanging wall as defined in Fig.~\ref{fig: 2-layer-sys}, fault dip angle ($f_\beta$), faulting depth ($z_f$), and maximum burial depth ($z_\text{max}$). The ranges and baseline values for these inputs are summarized in Table~\ref{tab:input-parameters}. This simplified system provides a controlled setting in which we can systematically explore the effects of lithological controls on fault permeability outcomes and perform sensitivity analysis using both local and global methods.

\begin{table}[!h]
    \centering
        \caption{Ranges and baseline values of lithological input parameters for the two-layer system.}
    \label{tab:input-parameters}
    \begin{tabular}{l|l|l}
        \toprule
        \textbf{Input Parameters} & \textbf{Range} &\textbf{Baseline}\\
        \hline
        Clay Fraction for Clay Layer $V_\text{cl}^c$ & $[0.5,0.9]$[-]& $0.7$\\
        \hline
        Clay Fraction for Sand Layer $V_\text{cl}^s$& $[0.1,0.3]$[-] &$0.2$\\
        \hline
        Fault Dip Angle $f_\beta$& $[50, 80]$[deg] & $65$[deg]\\
        \hline
        Faulting Depth $z_f$ & $[200,1000]$[m] & $600$[m]\\
        \hline
        Maximum Burial Depth $z_\text{max}$ & $[1000,3000]$[m] & $2000$[m]\\
        \botrule
    \end{tabular}
\end{table}

\section{Local Sensitivity Analysis}\label{sec: LSA}
To assess the individual influence of each lithological control on the predicted fault permeabilities, we begin with a local sensitivity analysis using the one-at-a-time (OAT) approach. In this method, each input parameter is varied independently across its specified range while keeping all other parameters fixed at their baseline values (as given in Table~\ref{tab:input-parameters}). This allows us to isolate the direct effect of each parameter on the output quantities of interest.

\begin{figure}[!h]
    \includegraphics[trim = 0 0 0 0, clip, width = \textwidth]{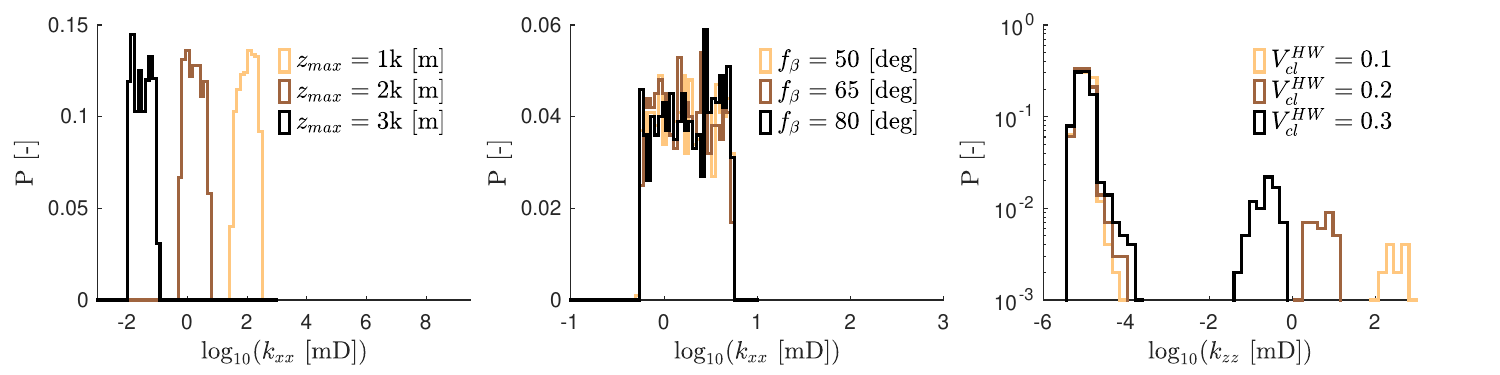}
\caption{Responses of permeability distributions to variations in individual parameters for the CSS scenario. Left: distribution changes of $\log_{10}(k_{xx})$ with varying $z_\text{max}$; Middle: distribution changes of $\log_{10}(k_{xx})$ with varying $f_\beta$; Right: distribution changes of $\log_{10}(k_{zz})$ with varying $V_\text{cl}^\text{HW}$.}
\label{fig: OAT}
\end{figure}

Fig.~\ref{fig: OAT} illustrates representative results for the CSS scenario, showing how variations in individual parameters---while holding all others fixed at their baseline values---affect the distribution of permeability in the dip-normal ($k_{xx}$) and dip-parallel ($k_{zz}$) directions. For example, increasing the maximum burial depth $z_\text{max}$ causes a clear shift in the $\log_{10}(k_{xx})$ distribution, while changes in fault dip $f_\beta$ produce comparatively small changes. In contrast, variation in $V_\text{cl}^\text{HW}$ leads to more complex changes in the shape of the $\log_{10}(k_{zz})$ distribution.

\begin{figure}[!h]
    \includegraphics[trim = 0 0 0 0, clip, width = 0.32\textwidth]{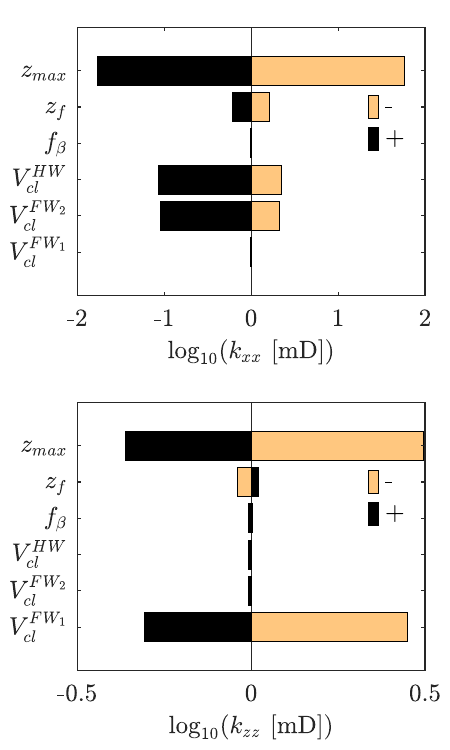}
        \includegraphics[trim = 0 0 0 0, clip, width = 0.32\textwidth]{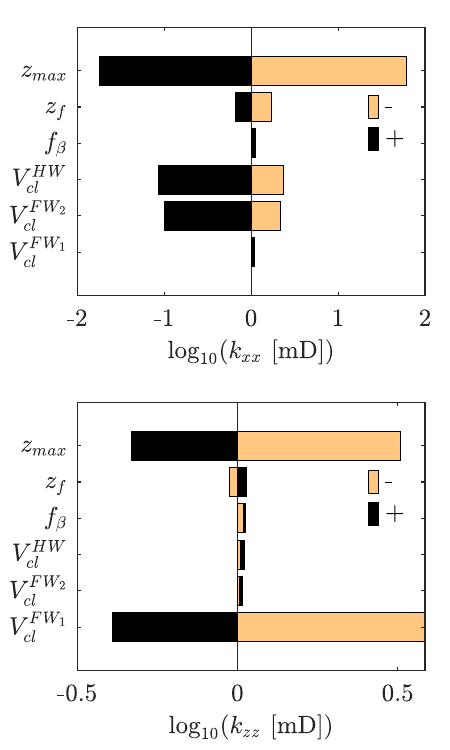}
            \includegraphics[trim = 0 0 0 0, clip, width = 0.32\textwidth]{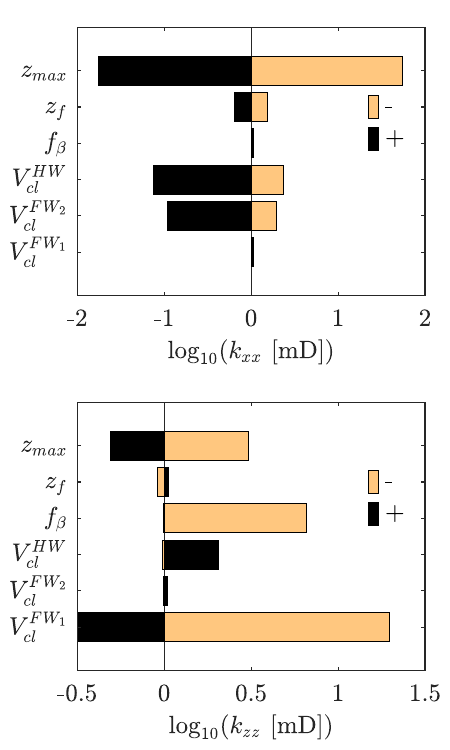}
    \caption{Tornado plots of the 10th (left), 50th (middle), and 90th (right) percentiles for $\log_{10}(k_{xx})$ (top row) and $\log_{10}(k_{zz})$ (bottom row) in the CSS scenario. Dark brown bars represent deviations caused by positive perturbations ($\Delta^+ Q^s_i$ in Eq.~\eqref{eq: tornado_plus}), while light orange bars represent deviations caused by negative perturbations ($\Delta^- Q^s_i$ in Eq.~\eqref{eq: tornado_minus}).}
    \label{fig: Tornado}
\end{figure}

To further quantify these effects in the output distributions, we generate tornado plots for the 10$^\text{th}$, 50$^\text{th}$, and 90$^\text{th}$ percentiles of $\log_{10}(k_{xx})$ and $\log_{10}(k_{zz})$, as shown in Fig.~\ref{fig: Tornado}. For each input parameter $p_i \in \{V_\text{cl}^{\text{FW}1}$, $V_\text{cl}^{\text{FW}2},V_\text{cl}^{\text{HW}},f_\beta, z_f, z_\text{max}\}$, we evaluate its influence on the fault permeability distributions by quantifying deviations in the $s^\text{th}$ percentile, denoted $Q^s$, when $p_i$ is perturbed positively and negatively from its baseline value, while all other parameters are held fixed at the baseline values. Specifically, the sensitivity measure for parameter $p_i$ is defined as
\begin{subequations}\label{eq: tornado}
\begin{align}
\Delta^+ Q^s_i &= Q^s(p_i^{\text{max}}) - Q^s(p_i^{\text{base}}), \label{eq: tornado_plus} \\
\Delta^- Q^s_i &= Q^s(p_i^{\text{min}}) - Q^s(p_i^{\text{base}}), \label{eq: tornado_minus}
\end{align}
\end{subequations}
where $p_i^\text{max}$ and $p_i^\text{min}$ represent the upper and lower bounds of parameter $p_i$ as given in Table~\ref{tab:input-parameters}, and $Q^s(p_i^\text{max/min})$ denotes the $s^\text{th}$ percentile of the permeability distribution computed by PREDICT under the corresponding perturbation. These deviations, $\Delta^+ Q^s_i$ (dark brown) and $\Delta^- Q^s_i$ (light orange), 
provide a directional measure of sensitivity and are used to construct the bars in the tornado plots shown in Fig.~\ref{fig: Tornado}. The length and asymmetry of each bar reflect both the magnitude and direction of each parameter’s influence on the specified output percentile.

The tornado plots in Fig.~\ref{fig: Tornado} reveal how different lithological parameters influence the statistical behavior of fault permeability in the CSS scenario. Across all percentiles, the maximum burial depth $z_\text{max}$ emerges as the most influential parameter, particularly for $k_{xx}$, where both positive and negative perturbations result in substantial shifts in the output distribution. This suggests that burial depth strongly influences fault material compaction, which in turn governs the reduction in permeability. In contrast, the fault dip $f_\beta$ exhibits minimal sensitivity across all percentiles and permeability directions, indicating that small variations in dip angle do not substantially alter the upscaled permeability in this structural setting. The clay fractions—especially $V_\text{cl}^{\text{FW}_1}$ and $V_\text{cl}^{\text{HW}}$—show moderate but asymmetric influence, suggesting nonlinear interactions between lithology and flow pathways. Notably, the impact of $V_\text{cl}^{\text{FW}_1}$ is more pronounced for $k_{zz}$ than for $k_{xx}$, as its location in the lower half of the footwall enables it to obstruct flow in the vertical direction while exerting little influence on horizontal connectivity. These trends are consistent across the 10th, 50th, and 90th percentiles, indicating that the dominant controls are robust across the tails and center of the distribution. Overall, the results demonstrate that while $z_\text{max}$ dominates the global scaling behavior, clay fraction variations contribute significantly to the anisotropy and spread of the permeability distribution. For completeness, we provide the results for the other three configurations in the supplementary material (Fig.~\ref{fig: Tornado-CSC}, \ref{fig: Tornado-SCC}, \ref{fig: Tornado-SCS}).

\section{Global Sensitivity Analysis}\label{sec: GSA}
While the local sensitivity analysis provides valuable insights into the individual effects of input parameters, it does not capture interactions between parameters or nonlinear dependencies that may arise when multiple variables vary simultaneously. The OAT approach assumes independence among inputs and may underestimate the influence of parameters that primarily contribute through synergistic effects or conditional dependencies. This limitation is particularly important in the context of the PREDICT modeling framework, where the sampling process involves conditional dependencies and correlation enforcement derived from physical constraints and empirical relationships—such as mechanical stability criteria and stratigraphic consistency. As a result, the input space explored by PREDICT is highly structured, and local sensitivity methods may fail to reveal important joint effects. To overcome these shortcomings and obtain a more comprehensive understanding of parameter influence, we now turn to a global sensitivity analysis, which explores the full input space and captures both main individual effects and total interaction effects. 

\subsection{Methodology and Computational Challenges}\label{sec: method}
Global sensitivity analysis (GSA) provides a quantitative framework for assessing how the uncertainty in model outputs ($Q$) can be attributed to uncertainties in the input parameters ($\mathbf p$), including both their individual contributions and interaction effects. In this study, we adopt a variance-based GSA approach, which decomposes the total output variance into contributions from each input parameter and their combinations. This allows us to quantitatively measure not only the main effect of each parameter but also how it interacts with others in driving the variability of the model response. A widely used metric in variance-based GSA is the Sobol’ sensitivity index~\cite{sobol1993sensitivity}, which includes two key components: the \textit{main-effect} Sobol’ index $S_i$, representing the fraction of output variance explained by variations in input parameter $p_i$ alone, and the \textit{total-effect} index $S_{T_i}$, which accounts for both the main effect and all higher-order interactions involving $p_i$. Together, these indices provide a comprehensive picture of the sensitivity landscape and enable us to distinguish between dominant, interacting, and negligible parameters in the model.

Typically, the Sobol' sensitivity indices are estimated via Monte Carlo simulation using ``fixing methods''~\cite{saltelli1993sensitivity}, which involve sampling the input space to approximate the conditional and marginal variances required for index computation. In our study, the input parameters are sampled from independent uniform distributions within the ranges specified in Table~\ref{tab:input-parameters}. The procedure for estimating the Sobol’ indices using this approach is outlined in Algorithm~\ref{alg: sobol_mc}. We refer the reader to~\cite{dellino2015uncertainty} for a comprehensive review of global sensitivity analysis.

\begin{algorithm}[!h]
\caption{Monte Carlo Estimation of Sobol' Sensitivity Indices via Fixing Method}
\label{alg: sobol_mc}
\begin{algorithmic}[1]
\State \textbf{Input:} Number of samples $N_{\text{mc}}$, input parameter space $\mathcal P\subset \mathbb R^{N_p}$ with prescribed density $\rho$, model~\eqref{eq: model}.
\State Sample two independent parameter sets from $\mathcal P$ with prescribed density $\rho$:
\begin{equation}
\{\mathbf{p}^{(k)}\}_{k = 1}^{N_\text{mc}} = \{(p_1^{(k)}, \ldots, p_{N_p}^{(k)})\}_{k = 1}^{N_\text{mc}} \mbox{ and } \{\hat{\mathbf{p}}^{(k)}\}_{k = 1}^{N_\text{mc}} = \{(\hat p_1^{(k)}, \ldots, \hat p_{N_p}^{(k)})\}_{k = 1}^{N_\text{mc}};
\end{equation}
 \State Evaluate model outputs $Q(\mathbf{p}^{(k)})$ and $Q(\hat{\mathbf{p}}^{(k)})$ from~\eqref{eq: model};
    \State Compute QoIs, e.g., the $s^\text{th}$ percentile of the output distributions $Q^s(\mathbf{p}^{(k)})$ and $Q^s(\hat{\mathbf{p}}^{(k)})$;
    \State Compute mean and total variance:
    \begin{equation}
    \begin{aligned}
        &E = \frac{1}{N_{\text{mc}}} \sum_{k=1}^{N_{\text{mc}}} Q^s(\mathbf p^{(k)}),\quad V = \frac{1}{N_{\text{mc}} - 1} \sum_{k=1}^{N_{\text{mc}}} \left(Q^s(\mathbf p^{(k)}) - E\right)^2\\
        &\hat E = \frac{1}{N_{\text{mc}}} \sum_{k=1}^{N_{\text{mc}}} Q^s(\hat{\mathbf p}^{(k)}),\quad \hat V = \frac{1}{N_{\text{mc}} - 1} \sum_{k=1}^{N_{\text{mc}}} \left(Q^s(\hat{\mathbf p}^{(k)}) - \hat E\right)^2
        \end{aligned}
        \end{equation}
\For{$i = 1$ to $N_p$}
    \State Construct a hybrid sample set $\{\mathbf h_i^{(k)}\}_{k=1}^{N_\text{mc}}$
    \begin{equation}
    \mathbf h_i^{(k)} = (p_1^{(k)}, \cdots, p_{i-1}^{(k)}, \hat p_i^{(k)}, p_{i+1}^{(k)}, \cdots, p_{N_p}^{(k)})
    \end{equation}
    \hspace{1em} by replacing the $i$-th element of $\mathbf p^{(k)}$ with the $i$-th element of $\hat{\mathbf{p}}^{(k)}$;
    \State Evaluate model outputs $Q(\mathbf{h}^{(i,k)})$ and compute QoIs $Q^s(\mathbf{h}^{(i,k)})$;
    \State Compute the main-effect and total-effect variances:
    \begin{equation}\label{eq:sobol}
        \begin{aligned}
    &V_i = \frac{2N_{\text{mc}}}{2N_{\text{mc}} - 1} \left( \frac{1}{N_{\text{mc}}} \sum_{k=1}^{N_{\text{mc}}} Q^s(\mathbf p^{(k)}) Q^s(\mathbf h^{(i,k)}) - \frac{E^2 + \hat{E}^2}{2} + \frac{V + \hat{V}}{4} \right)\\
   &T_i = \frac{1}{2N_{\text{mc}}} \sum_{k=1}^{N_{\text{mc}}} \left(Q^s(\hat{\mathbf p}^{(k)}) - Q^s(\mathbf h^{(i, k)})\right)^2
   \end{aligned}
   \end{equation}
    \State Compute Sobol' main-effect and total-effect indices:
\begin{equation}
    S_i = V_i / V,\quad S_{T_i} = T_i / V;
\end{equation}
\EndFor
\State \textbf{Output:} Sobol' indices $S_i$ and $S_{T_i}$ for $i = 1, \ldots, N_p$.
\end{algorithmic}
\end{algorithm}

While this approach is straightforward to implement and widely used, it is computationally intensive—particularly in high-dimensional and complex systems. The method requires $(N_p + 2)N_\text{mc}$ model evaluations, where $N_\text{mc}$ is the number of Monte Carlo samples and grows rapidly with the number of input parameters $N_p$. This poses a significant challenge for our application, as each model evaluation involves a full realization of the PREDICT workflow, which includes $N_\text{sim}$ repeated simulations comprising conditional sampling, physically constrained material placement, and flow-based upscaling. Among these steps, the flow-based upscaling—performed by MRST on a $100 \times 100$ mesh—is the most computationally demanding, accounting for approximately 80\% of the runtime per simulation. To accurately estimate the permeability distribution from a single PREDICT model evaluation, $N_\text{sim}$ is typically on the order of 1,000. For the case of $N_p = 6$, $N_\text{mc} = 5{,}000$, and $N_\text{sim} = 1{,}000$, this results in $40{,}000{,}000$ flow-based upscaling operations, rendering direct Monte Carlo-based sensitivity estimation prohibitively expensive.

\subsection{A Neural Network-based Surrogate Model for Flow-based Upscaling}
\begin{figure}[!h]
    \includegraphics[trim = 25 90 120 190, clip, width = \textwidth]{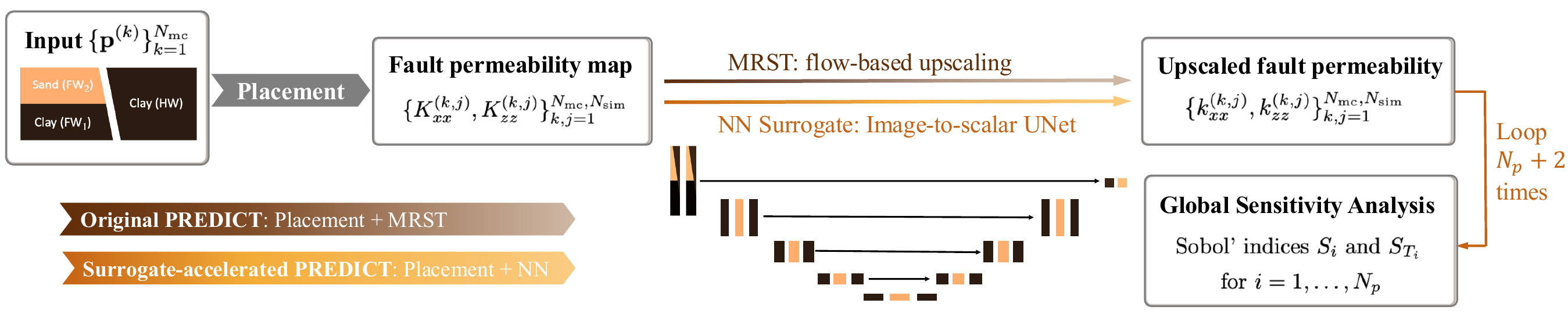}
    \caption{Proposed workflow to enable global sensitivity analysis: replacing flow-based MRST upscaling by a CNN surrogate within PREDICT.}
    \label{fig: GSA-workflow}
\end{figure}

As we have identified, the majority of the computational bottleneck in performing global sensitivity analysis arises from the repeated execution of flow-based upscaling. To alleviate this cost, we propose to develop a neural network-based surrogate model that emulates the flow-based upscaling step within the PREDICT workflow (see Fig.~\ref{fig: GSA-workflow}). By replacing the physics-based MRST solver with a computationally efficient yet accurate surrogate, we aim to significantly reduce the overall computational burden while maintaining sufficient accuracy for downstream uncertainty quantification tasks.

To train the surrogate model, we construct a dataset of input–output pairs, where each input is a fault permeability map $(K_{xx}, K_{zz})$ defined on a $100 \times 100$ grid. These maps are generated from the material placement step within PREDICT, and the corresponding outputs are pairs of upscaled scalar permeability values $(k_{xx}, k_{zz})$ computed using MRST. The training dataset is built by sampling $N_\text{mc}^{\text{train},s} = 100$ parameter combinations within the range in Table~\ref{tab:input-parameters} for each of the $s = $ CSC, CSS, SCC, SCS geological scenarios shown in Fig.~\ref{fig: 2-layer-sys}. For each parameter input, we run $N_\text{sim}^\text{train} = 100$ MRST simulations, yielding a total of 40,000 input–output pairs for training. We note that $N_\text{sim}^\text{train} = 100$ is typically insufficient to accurately characterize the output distribution of permeability for a single stratigraphy. Furthermore, using only $N_\text{mc}^{\text{train},s} = 100$ parameter samples per scenario is inadequate for reliable computation of Sobol’ indices. These limitations highlight the need for a computationally efficient and accurate surrogate model.

We employ a convolutional neural network (CNN) architecture, UNet, tailored for image-to-scalar regression~\cite{ronneberger2015u}. The network takes the two 2D permeability fields $(K_{xx}, K_{zz})$ as input and outputs two scalar values representing $k_{xx}$ and $k_{zz}$, i.e.,
\begin{equation}
    (k_{xx},k_{zz})\approx \mathcal N_\Theta(K_{xx}, K_{zz}),
\end{equation}
where $\Theta$ is obtained by minimizing a mean absolute error (MAE) loss between the predicted and true permeability values. We choose MAE for its robustness to outliers and its tendency to yield more balanced errors across the range of permeability values; in our experiments, it also outperforms mean squared error (MSE) in terms of prediction accuracy measured by~\eqref{eq: wasserstein}. The architecture of UNet and the training code with specified hyperparameters is provided in Github (\hl{}).

\begin{figure}[!h]
    \includegraphics[trim = 0 0 0 0, clip, width = \textwidth]{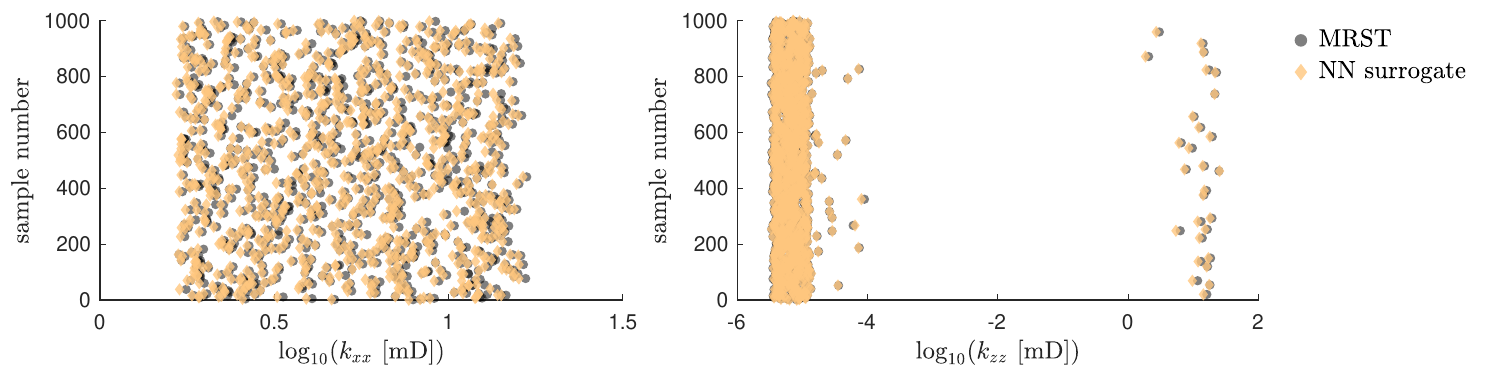}
    \includegraphics[trim = 0 0 0 0, clip, width = \textwidth]{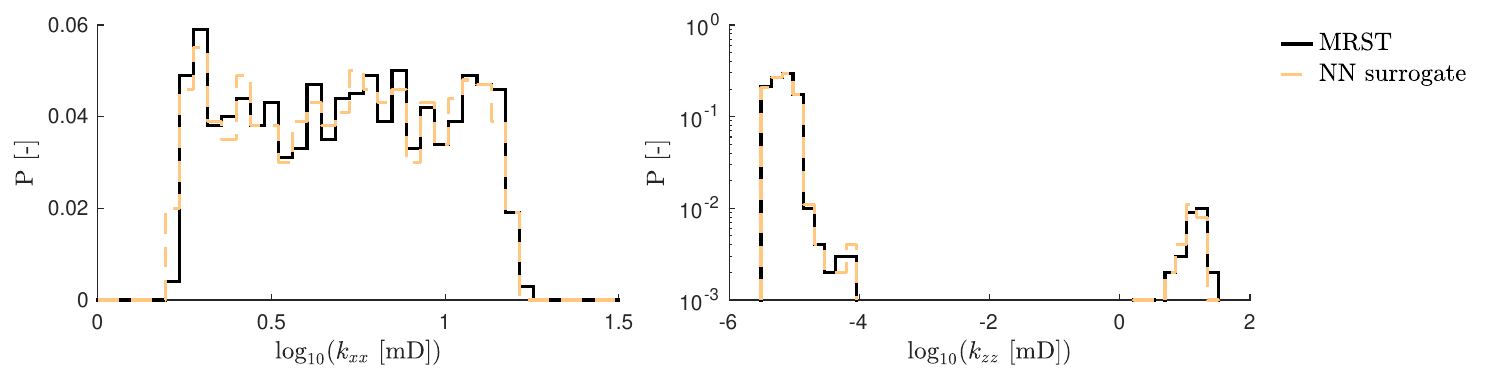}
    \caption{Validation of the surrogate model for a representative test input in the CSS scenario. \textbf{Top row:} Sample-wise comparison of predicted (NN surrogate) and reference (MRST) permeability values for $\log_{10}(k_{xx})$ (left) and $\log_{10}(k_{zz})$ (right). \textbf{Bottom row:} Probability distributions of predicted and reference values for $\log_{10}(k_{xx})$ (left) and $\log_{10}(k_{zz})$ (right).}
    \label{fig: val-results}
\end{figure}

After training, we validate the accuracy of the surrogate model using a representative task within the PREDICT workflow. Specifically, for a randomly sampled parameter combination $\mathbf{p}^* \notin \{\mathbf{p}^{(k)}\}_{k=1}^{N_\text{mc}^{\text{train},s}}$—unseen during training—from one of the four geological scenarios, we generate $N_\text{sim}^{\text{val},s} = 1{,}000$ realizations of permeability fields and validate the performance of the neural network surrogate for upscaling, using the MRST simulation as the reference. We assess the model’s performance by comparing both the sample-wise prediction errors and the resulting output distribution against those obtained from the same number of high-fidelity MRST upscaling simulations. Fig.~\ref{fig: val-results} (top row) shows that the NN-predicted values closely follow the MRST results across all samples, with only minor deviations, indicating that the surrogate model preserves pointwise accuracy even for inputs outside the training set. The bottom row presents probability histograms of the surrogate and MRST outputs. The close alignment between the two distributions confirms that the surrogate not only achieves sample-wise fidelity but also accurately reproduces the statistical structure of the output. Notably, the dip-parallel permeability ($k_{zz}$) in this CSS scenario, which often exhibits greater variability and a bimodal shape with a small tail, is also well-approximated. These results demonstrate the surrogate’s effectiveness in capturing both pointwise and distributional characteristics of the upscaled permeability.

The validation cases in the other three geological scenarios yield similarly accurate results. We summarize the quantitative performance of the surrogate model in Table~\ref{tab: val-results}, using the Wasserstein distance~\cite{panaretos2019} between the predicted and reference permeability distributions as the evaluation metric. For two one-dimensional distribution $P$ and $\tilde P$ with cumulative distribution functions (CDFs) $F_P(x)$ and $F_{\tilde P}(x)$, the Wasserstein distance (alson known as the Earth Mover's Distance) is defined as
\begin{equation}\label{eq: wasserstein-cont}
    \mathcal W_1(P,\tilde P) = \int_0^1|F_P^{-1}(u) -F^{-1}_{\tilde P}(u)|du,
\end{equation}
where $F_P^{-1}(u)$ and $F^{-1}_{\tilde P}(u)$ are the inverse CDFs. This integral computes the average horizontal distance between the two quantile curves across all quantile levels $u\in[0,1]$. For the empirical comparison based on $N_\text{sim}^{\text{val},s} = 1{,}000$ realizations, the Wasserstein distance between the empirical distributions of surrogate outputs and high-fidelity MRST results is computed as
\begin{equation}\label{eq: wasserstein}
\mathcal W(\kappa, \tilde\kappa) = \frac{1}{N_\text{sim}^{\text{val},s}}\sum_{k = 1}^{N_\text{sim}^{\text{val},s}}|\kappa^{(k)}-\tilde{\kappa}^{(k)}|,
\end{equation}
where $\tilde \kappa^{(k)}$ and $\kappa^{(k)}$ denote the $k$-th sorted samples from the NN surrogate and MRST permeability distributions, respectively. This formulation corresponds to the Wasserstein distance between two one-dimensional empirical distributions, which reduces to the $L^1$ distance between their sorted samples. Each index $k$ corresponds to the empirical quantile level $u_k = k/N_\text{sim}^{\text{sim},s}$, and thus the inverse CDFs can be approximated as
\begin{equation}
    F_P^{-1}(u_k)\approx \kappa^{(k)},\qquad F_{\tilde P}^{-1}(u_k)\approx \tilde\kappa^{(k)}.
\end{equation}
Therefore, \eqref{eq: wasserstein} is obtained by replacing the integral in~\eqref{eq: wasserstein-cont} with a Riemann sum over uniformly spaced quantile levels.

\begin{table}[!h]
    \centering
        \caption{Wasserstein distance between the NN predicted permeability distribution and the MRST reference distribution, evaluated over $N_\text{sim}^{\text{val},s} = 1{,}000$ samples. Results are reported for four representative cases from four geological scenarios (CSC, CSS, SCC, SCS).}
    \label{tab: val-results}
    \begin{tabular}{l|l|l|l|l}
        \toprule
        & \textbf{CSC} &\textbf{CSS} & \textbf{SCC} &\textbf{SCS}\\
        \hline
        $\mathcal W(\log_{10}(k_{xx}),\log_{10}(\tilde k_{xx}))$[mD] &0.0030&0.0086&0.0016&0.0034 \\
        \hline        $\mathcal W(\log_{10}(k_{zz}),\log_{10}(\tilde k_{zz}))$[mD] & 0.0013&0.0034&0.0040&0.0027\\
        \botrule
    \end{tabular}
\end{table}

While the absolute values of $\mathcal W$ are small in Table~\ref{tab: val-results}, their magnitudes are better interpreted when normalized by the spread of the corresponding MRST reference distributions. Across all four geological scenarios, the normalized Wasserstein distance $\mathcal W / \sigma_{\text{MRST}}$ remains below $3\%$, where $\sigma_{\text{MRST}}$ denotes the standard deviation of the high-fidelity permeability distributions. This indicates that the surrogate reproduces the full permeability distribution with errors that are very small relative to its intrinsic variability, capturing not only mean trends but also higher-order statistical characteristics.

This set of $4{,}000$ input–output pairs is also used as the validation dataset during training. Training the NN surrogate takes approximately 23 seconds per epoch on a single NVIDIA A100 Tensor Core GPU, and both the training and validation losses plateau in fewer than 1,000 epochs. After training, the inference time for each upscaling operation is approximately 0.39 milliseconds, whereas each MRST flow-based upscaling takes approximately $26$ seconds when run on a Mac Pro with an Apple M2 Pro chip. Since the material placement step accounts for only a small fraction of the total upscaling time in the PREDICT framework (approximately $8.5$ seconds), replacing the MRST-based upscaling with the NN surrogate results in more than a 3,000-fold acceleration per run of PREDICT.

\subsection{Results and Discussion}
The efficient yet accurate surrogate model allows us to accelerate the $40{,}000{,}000$ upscaling operations required for global sensitivity analysis in each scenario (as discussed in Section~\ref{sec: method}), which would otherwise be computationally prohibitive due to the cost of MRST simulations. We compute the Sobol’ indices using Algorithm~\ref{alg: sobol_mc}. In the input–output model~\eqref{eq: model}, we replace the MRST-based flow upscaling with the NN surrogate, enabling the analysis to be computationally feasible.

\begin{figure}[!h]
    \includegraphics[trim = 0 0 0 0, clip, width = 0.32\textwidth]{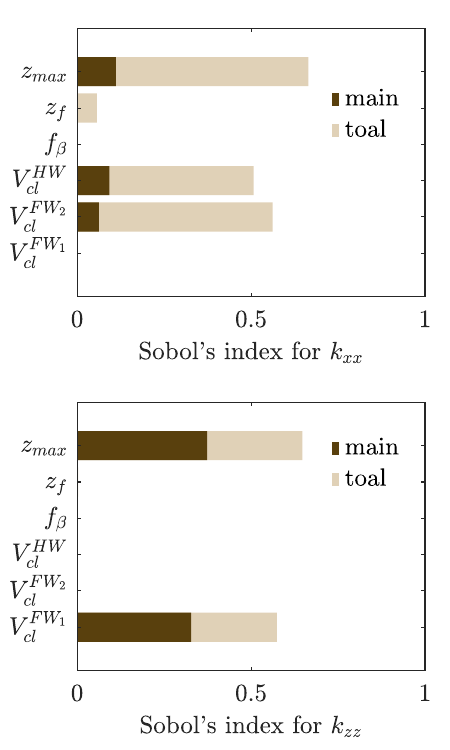}
        \includegraphics[trim = 0 0 0 0, clip, width = 0.32\textwidth]{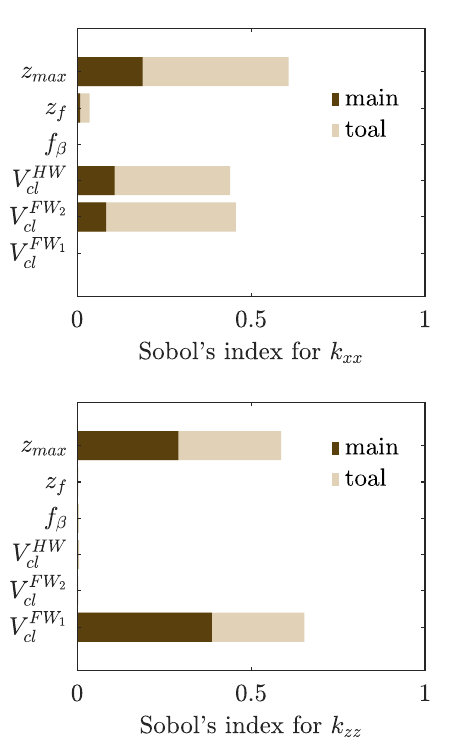}
            \includegraphics[trim = 0 0 0 0, clip, width = 0.32\textwidth]{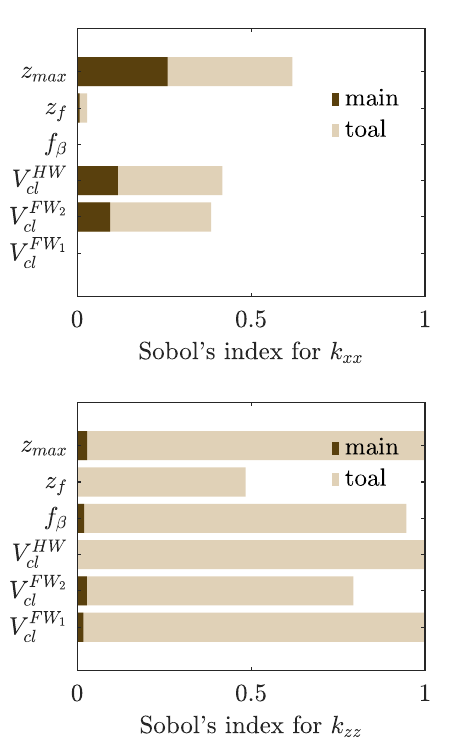}
    \caption{Sobol' indices of the 10th (left), 50th (middle), and 90th (right) percentiles for $k_{xx}$ (top row) and $k_{zz}$ (bottom row) in the CSS scenario using $N_\text{mc} = 5{,}000$ and $N_\text{sim} = 1000$. Dark brown bars represent main-effect index $S_i$, while light orange bars represent total-effect index $S_{T_i}$ calculated by Eq.~\eqref{eq:sobol}.}
    \label{fig: Sobol-CSS}
\end{figure}

Figure~\ref{fig: Sobol-CSS} presents the Sobol’ sensitivity indices for the 10th, 50th, and 90th percentiles of the fault permeability distributions in the CSS scenario, computed using the surrogate-accelerated workflow with $N_\text{mc} = 5{,}000$ and $N_\text{sim} = 1{,}000$. The top row shows results for $k_{xx}$ (dip-normal permeability), and the bottom row shows results for $k_{zz}$ (dip-parallel permeability). For each output percentile, we plot both the main-effect index $S_i$ (dark brown) and the total-effect index $S_{T_i}$ (light orange) for all six lithological input parameters. 

The results highlight several important trends. First, consistent with the local analysis in Fig.~\ref{fig: Tornado}, the maximum burial depth $z_{\text{max}}$ emerges as the most influential parameter across all percentiles and both permeability directions. Its high total-effect indices indicate not only a strong individual contribution but also significant interactions with other parameters, especially for $k_{xx}$. This suggests that burial depth exerts a dominant control on fault material compaction and overall permeability reduction, with its impact further modulated by the values of other inputs.

Second, the clay volume fraction in the hanging wall, $V^{\text{HW}}_{\text{cl}}$, exhibits a moderate influence on $k_{xx}$, with a large portion of its contribution arising from interaction effects rather than its individual impact. For $k_{zz}$, its effect is negligible in the 10th and 50th percentiles but becomes significant in the 90th percentile due to strong interaction effects—likely reflecting its role in controlling the high permeability mode when clay smear from FW$_1$ is not plugging the fault. This explains the intricate distribution shift observed in the right column of Fig.~\ref{fig: OAT}. More broadly, the pronounced total-effect indices across all input parameters in the 90th percentile of $k_{zz}$ suggest that the upper tail of the $k_{zz}$ distribution is shaped by intricate, nonlinear interactions among all lithological controls.

Third, the clay fractions in the footwall layers, $V^{\text{FW1}}_{\text{cl}}$ and $V^{\text{FW2}}_{\text{cl}}$, show differentiated influence on the two permeability components: $V^{\text{FW2}}_{\text{cl}}$ contributes moderately to $k_{xx}$, while $V^{\text{FW1}}_{\text{cl}}$ has a comparable effect on $k_{zz}$. In both cases, their sensitivity is dominated by interaction effects rather than main effects, indicating that their influence is highly dependent on the values of other input parameters. FW$_1$ is responsible for generating low-permeability smears that act as barriers to flow.

In contrast, the fault dip angle $f_\beta$ consistently shows low sensitivity across all percentiles and both directions, reaffirming the finding from the local analysis that moderate changes in structural orientation have a negligible effect on effective fault permeability in this setting. The faulting depth $z_f$ shows modest influence, with a slightly higher impact on $k_{zz}$, especially at the upper percentile. This may reflect its interaction with stratigraphic layering in determining the dip-parallel permeability depending on the value of $V_\text{cl}^{\text{FW}_1}$.

In addition to the CSS case, we also analyzed the CSC, SCC, and SCS scenarios; the detailed results are shown in Figs.~\ref{fig: Tornado-CSC}–\ref{fig: Sobol-SCS} in the Appendix.
\begin{itemize}
\item The CSC scenario (Figs.~\ref{fig: Tornado-CSC}–\ref{fig: Sobol-CSC}) confirms that burial depth $z_{\text{max}}$ remains a dominant control on both $k_{xx}$ and $k_{zz}$, as indicated by the tornado plots. Additionally, $V^{\text{HW}}_{\text{cl}}$ is significantly more important for $k_{xx}$ than in the CSS scenario, because it controls clay smear sealing across the fault. The global sensitivity analysis reveals that these influences are further amplified by strong interaction effects, particularly in $z_{\text{max}}$ in the upper percentiles of $k_{zz}$. In addition, $V^{\text{FW1}}_{\text{cl}}$ shows a clear impact on $k_{zz}$ at high percentiles, consistent with the local analysis where its main effect is evident. The Sobol' indices, however, highlight that this contribution is not purely additive but also involves substantial interaction effects, underscoring the role of footwall smearing in reducing vertical permeability. This is a relevant finding, given that fault sealing typically relies on hangingwall smearing only~\cite{yielding1997}.

\item The SCC scenario (Figs.~\ref{fig: Tornado-SCC}–\ref{fig: Sobol-SCC}) again identifies burial depth $z_{\text{max}}$ as the dominant control on both permeability components. Global sensitivity analysis clarifies that, beyond this primary effect, $V^{\text{HW}}_{\text{cl}}$ and $V^{\text{FW2}}_{\text{cl}}$ contribute moderately and mainly through interactions, particularly for $k_{xx}$; this is because both of these layers can contribute clay smear throughout the throw window in this configuration. In contrast, $V^{\text{FW1}}_{\text{cl}}$ plays only a minor role with much smaller indices than those of the clay-rich layers. Both analyses show negligible sensitivity to $z_f$ and $f_\beta$ for the ranges considered. Overall, the global (Sobol') results reveal an interaction-driven sharing of influence between $V^{\text{HW}}_{\text{cl}}$ and $V^{\text{FW2}}_{\text{cl}}$ that the local tornado plots understate, while confirming the limited importance of $V^{\text{FW1}}_{\text{cl}}$, $z_f$, and $f_\beta$.

\item The SCS scenario (Figs.~\ref{fig: Tornado-SCS}–\ref{fig: Sobol-SCS}) shows that the upper tail of $k_{zz}$ is highly interaction–driven: the Sobol' total-effect indices at the 90th percentile are large across \emph{most} inputs and substantially exceed the corresponding main effects, indicating that percentile-extreme outcomes arise from coupled effects rather than any single parameter alone. Consistent between local (tornado) and global (Sobol') analyses, burial depth $z_{\text{max}}$ remains the most influential parameter overall, and $V^{\text{FW2}}_{\text{cl}}$ makes a clear, moderate contribution—particularly to $k_{xx}$. The global analysis further reveals that $V^{\text{FW1}}_{\text{cl}}$ and $V^{\text{HW}}_{\text{cl}}$ materially shape the high-percentile behavior of $k_{zz}$ through interactions, an effect that the tornado plots understate. Parameters $z_f$ and $f_\beta$ are negligible across percentiles for both $k_{xx}$ and $k_{zz}$.
\end{itemize}

Overall, the global sensitivity analysis reveals complex and nonlinear interactions among lithological parameters, which are not captured by local methods. The comparison of main- and total-effect indices highlights the importance of considering both individual and interactive contributions, especially for parameters like $V^{\text{HW}}_{\text{cl}}$ and $z_f$. These insights demonstrate the power of the surrogate-accelerated framework for quantifying uncertainty and identifying dominant and/or neglible controls in geologic systems.

\section{Conclusion and Future Work}\label{sec: conclusion}

In this study, we explored the sensitivity of fault permeability to lithological controls using both local and global methods within the PREDICT framework. The OAT local sensitivity analysis offers a simple and intuitive means to assess the effect of individual input parameters by varying them independently around a fixed baseline. However, it suffers from several limitations: it does not capture any nonlinear dependencies or interaction effects, its results are highly sensitive to the choice of baseline values, and it lacks a quantitative measure of parameter importance across the full input space. In contrast, variance-based global sensitivity analysis (GSA) using Sobol’ indices provides a rigorous and comprehensive assessment of input-output relationships. It quantifies both the main effect of each parameter and the total effect that includes all higher-order interactions, offering insights into the full distributional behavior of the outputs. This makes GSA particularly valuable for systems like PREDICT, where the outputs may exhibit bimodal or heavy-tailed distributions due to complex, conditionally dependent geological processes. However, the application of GSA is computationally prohibitive when relying on high-fidelity MRST-based flow simulations. To overcome this bottleneck, we developed a convolutional neural network surrogate model based on the UNet architecture to emulate the upscaling step. The surrogate achieved high prediction accuracy and enabled tractable global sensitivity analysis over millions of model evaluations. This surrogate-accelerated GSA revealed strong interaction effects and intricate dependencies among input parameters—insights that are critical for understanding fault behavior and would remain inaccessible through local methods alone.

Future work will focus on extending the surrogate modeling approach to three-dimensional fault geometries and incorporating more complex stratigraphic structures. This will require the introduction of new parameterizations to describe stratigraphy and the development of generalizable surrogates that can adapt across different geological configurations. An important direction is to quantify the sensitivity of parameters that govern stratigraphic complexity and assess how much detail is feasible for a given throw window. The insights from global sensitivity analysis can also be used to interpret field observations, by identifying negligible parameters and accounting for key interactions. Ultimately, the knowledge gained from this study can be translated to full-field reservoir modeling, supporting more robust uncertainty quantification in applications such as large-scale geologic carbon storage.

\clearpage

\appendix
\paragraph{\textsc{Appendix}}
\renewcommand{\thefigure}{A\arabic{figure}} 
\renewcommand{\thetable}{A\arabic{table}}   
\setcounter{figure}{0}                      
\setcounter{table}{0}

\begin{figure}[!h]
    \includegraphics[trim = 0 0 0 0, clip, width = 0.32\textwidth]{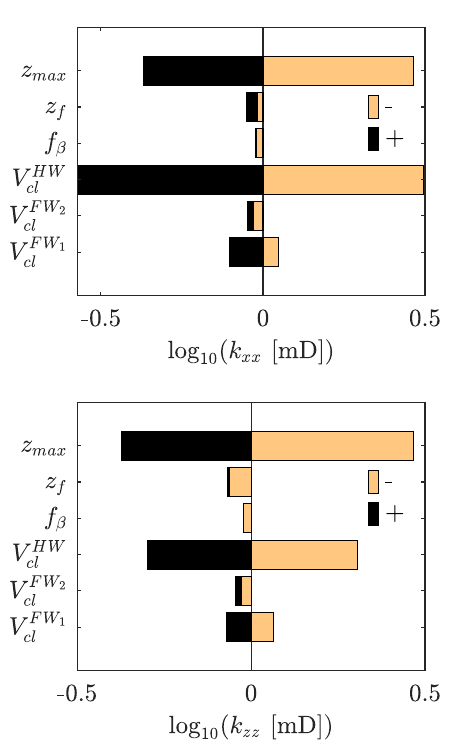}
        \includegraphics[trim = 0 0 0 0, clip, width = 0.32\textwidth]{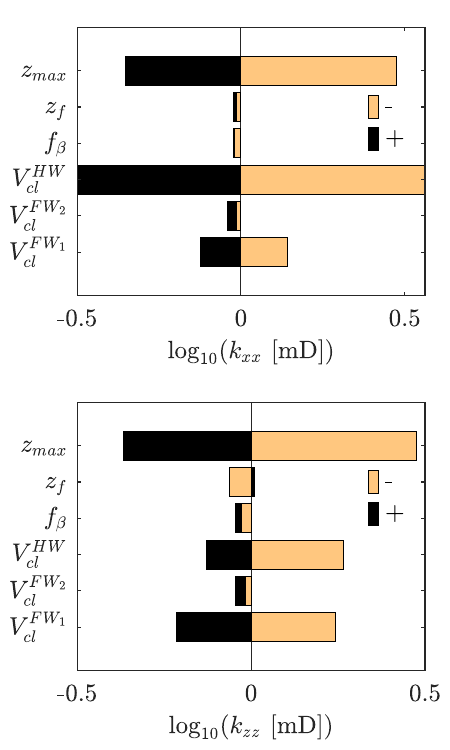}
            \includegraphics[trim = 0 0 0 0, clip, width = 0.32\textwidth]{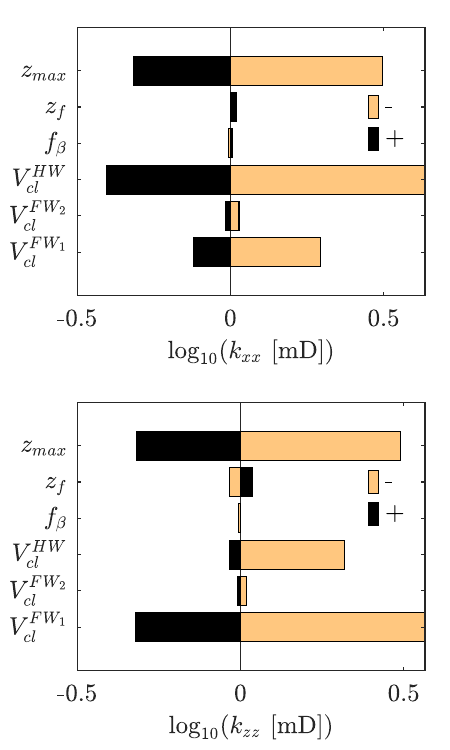}
    \caption{Tornado plots of the 10th (left), 50th (middle), and 90th (right) percentiles for $\log_{10}(k_{xx})$ (top row) and $\log_{10}(k_{zz})$ (bottom row) in the CSC scenario. Dark brown bars represent deviations caused by positive perturbations ($\Delta^+ Q^s_i$ in Eq.~\eqref{eq: tornado_plus}), while light orange bars represent deviations caused by negative perturbations ($\Delta^- Q^s_i$ in Eq.~\eqref{eq: tornado_minus}).}
    \label{fig: Tornado-CSC}
\end{figure}

\begin{figure}[!h]
    \includegraphics[trim = 0 0 0 0, clip, width = 0.32\textwidth]{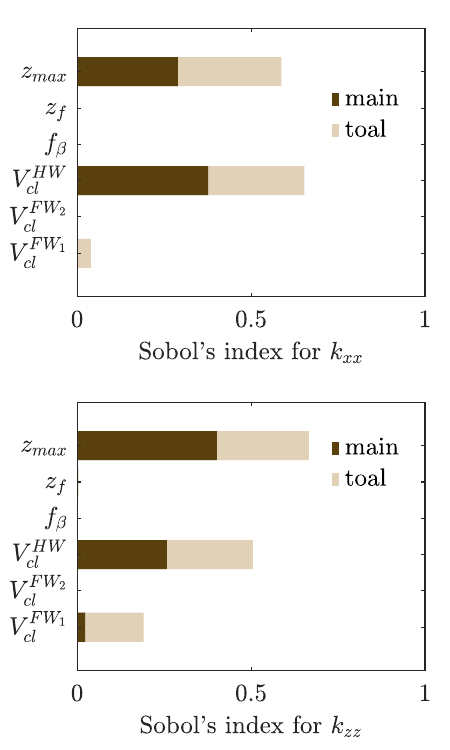}
        \includegraphics[trim = 0 0 0 0, clip, width = 0.32\textwidth]{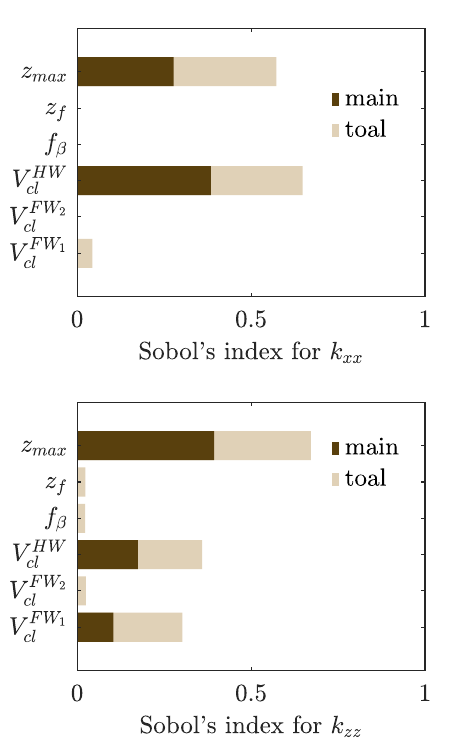}
            \includegraphics[trim = 0 0 0 0, clip, width = 0.32\textwidth]{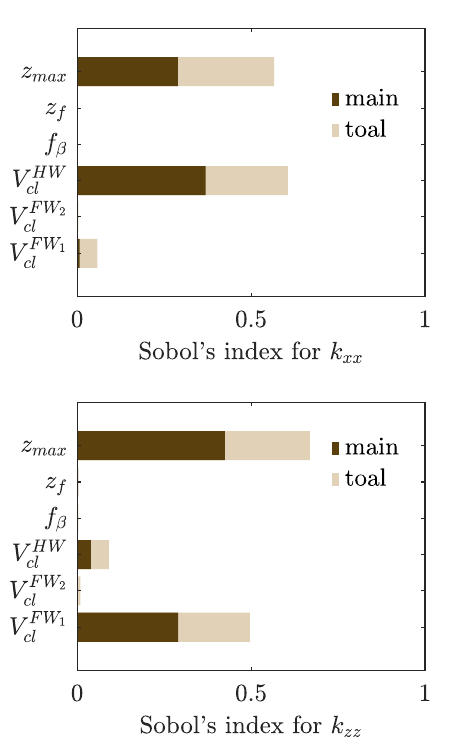}
    \caption{Sobol' indices of the 10th (left), 50th (middle), and 90th (right) percentiles for $k_{xx}$ (top row) and $k_{zz}$ (bottom row) in the CSC scenario using $N_\text{mc} = 5{,}000$ and $N_\text{sim} = 1000$. Dark brown bars represent main-effect index $S_i$, while light orange bars represent total-effect index $S_{T_i}$ calculated by Eq.~\eqref{eq:sobol}.}
    \label{fig: Sobol-CSC}
\end{figure}

\begin{figure}[!h]
    \includegraphics[trim = 0 0 0 0, clip, width = 0.32\textwidth]{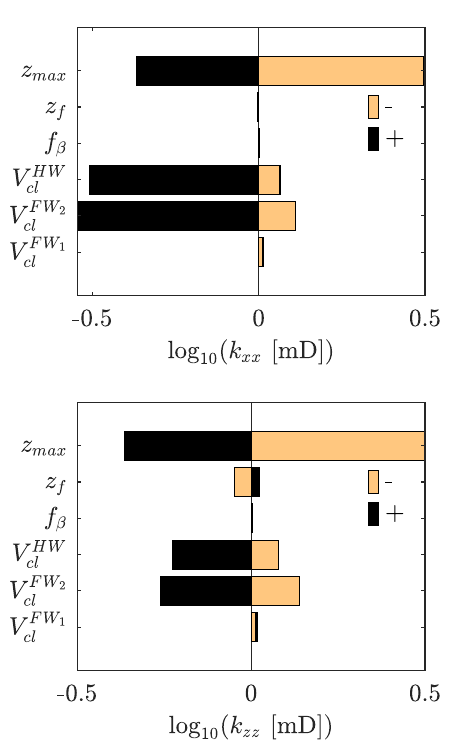}
        \includegraphics[trim = 0 0 0 0, clip, width = 0.32\textwidth]{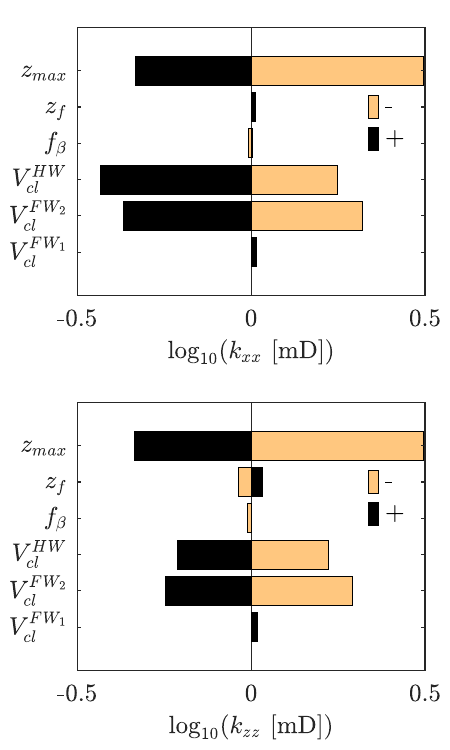}
            \includegraphics[trim = 0 0 0 0, clip, width = 0.32\textwidth]{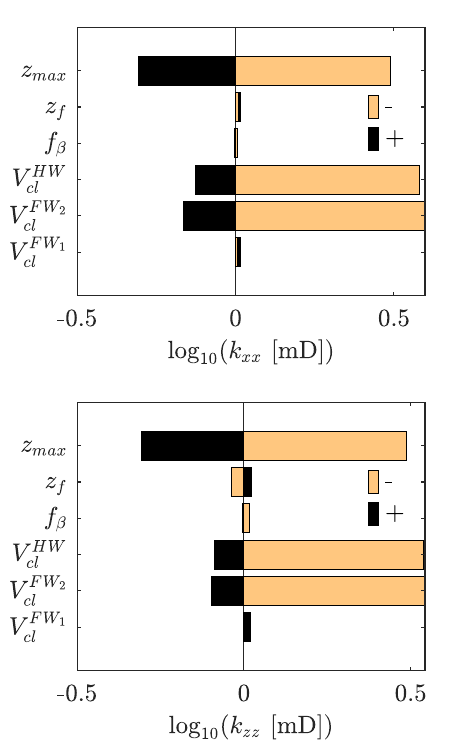}
    \caption{Tornado plots of the 10th (left), 50th (middle), and 90th (right) percentiles for $\log_{10}(k_{xx})$ (top row) and $\log_{10}(k_{zz})$ (bottom row) in the SCC scenario. Dark brown bars represent deviations caused by positive perturbations ($\Delta^+ Q^s_i$ in Eq.~\eqref{eq: tornado_plus}), while light orange bars represent deviations caused by negative perturbations ($\Delta^- Q^s_i$ in Eq.~\eqref{eq: tornado_minus}).}
    \label{fig: Tornado-SCC}
\end{figure}

\begin{figure}[!h]
    \includegraphics[trim = 0 0 0 0, clip, width = 0.32\textwidth]{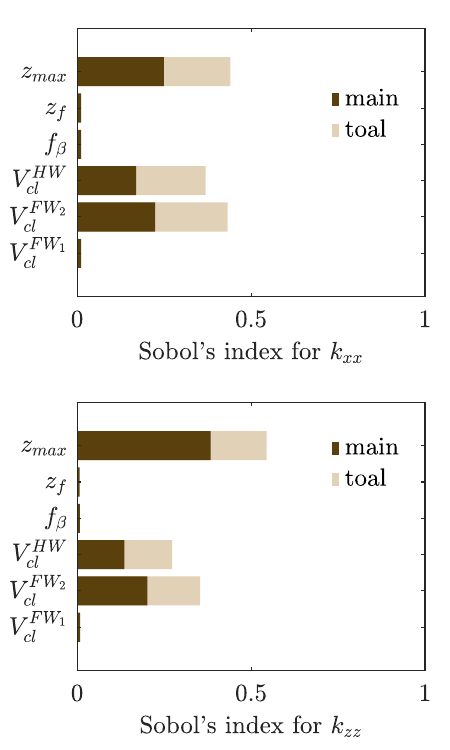}
        \includegraphics[trim = 0 0 0 0, clip, width = 0.32\textwidth]{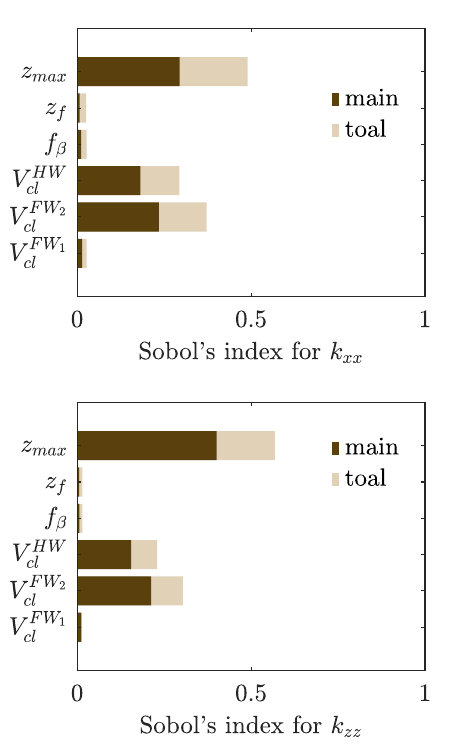}
            \includegraphics[trim = 0 0 0 0, clip, width = 0.32\textwidth]{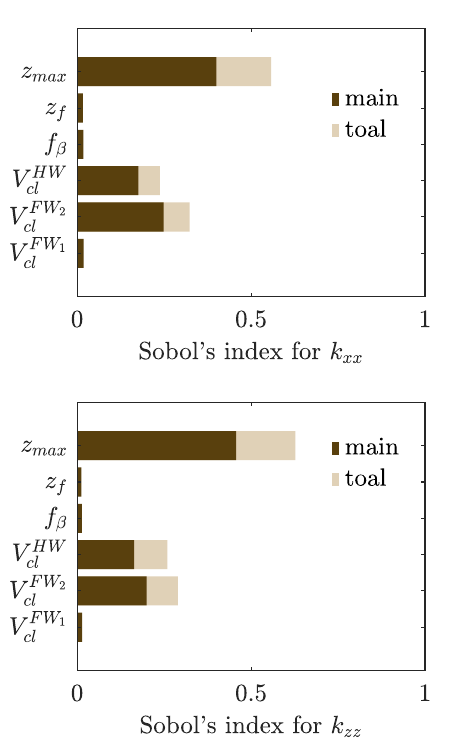}
    \caption{Sobol' indices of the 10th (left), 50th (middle), and 90th (right) percentiles for $k_{xx}$ (top row) and $k_{zz}$ (bottom row) in the SCC scenario using $N_\text{mc} = 5{,}000$ and $N_\text{sim} = 1000$. Dark brown bars represent main-effect index $S_i$, while light orange bars represent total-effect index $S_{T_i}$ calculated by Eq.~\eqref{eq:sobol}.}
    \label{fig: Sobol-SCC}
\end{figure}

\begin{figure}[!h]
    \includegraphics[trim = 0 0 0 0, clip, width = 0.32\textwidth]{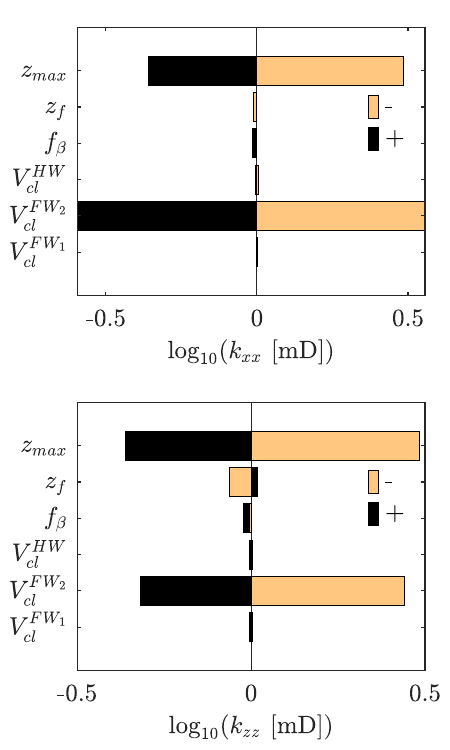}
        \includegraphics[trim = 0 0 0 0, clip, width = 0.32\textwidth]{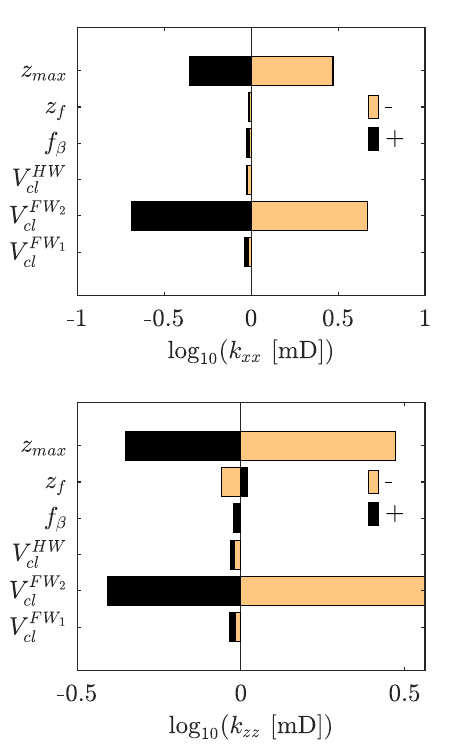}
            \includegraphics[trim = 0 0 0 0, clip, width = 0.32\textwidth]{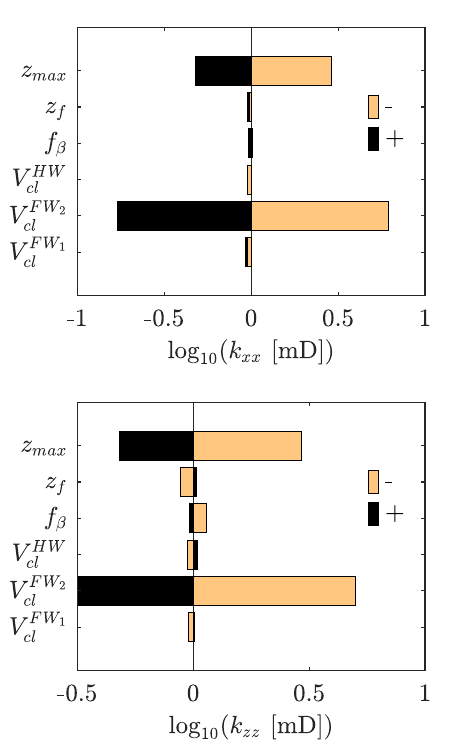}
    \caption{Tornado plots of the 10th (left), 50th (middle), and 90th (right) percentiles for $\log_{10}(k_{xx})$ (top row) and $\log_{10}(k_{zz})$ (bottom row) in the SCS scenario. Dark brown bars represent deviations caused by positive perturbations ($\Delta^+ Q^s_i$ in Eq.~\eqref{eq: tornado_plus}), while light orange bars represent deviations caused by negative perturbations ($\Delta^- Q^s_i$ in Eq.~\eqref{eq: tornado_minus}).}
    \label{fig: Tornado-SCS}
\end{figure}

\begin{figure}[!h]
    \includegraphics[trim = 0 0 0 0, clip, width = 0.32\textwidth]{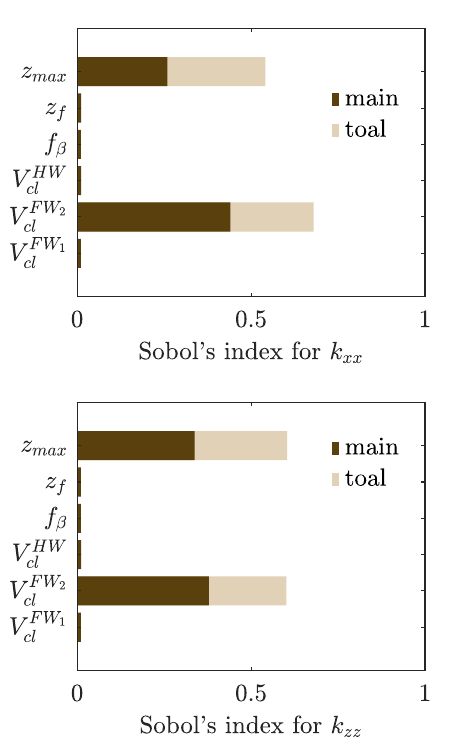}
        \includegraphics[trim = 0 0 0 0, clip, width = 0.32\textwidth]{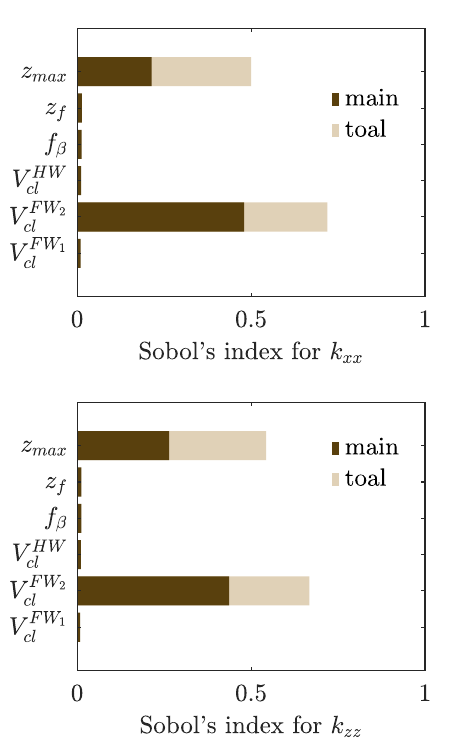}
            \includegraphics[trim = 0 0 0 0, clip, width = 0.32\textwidth]{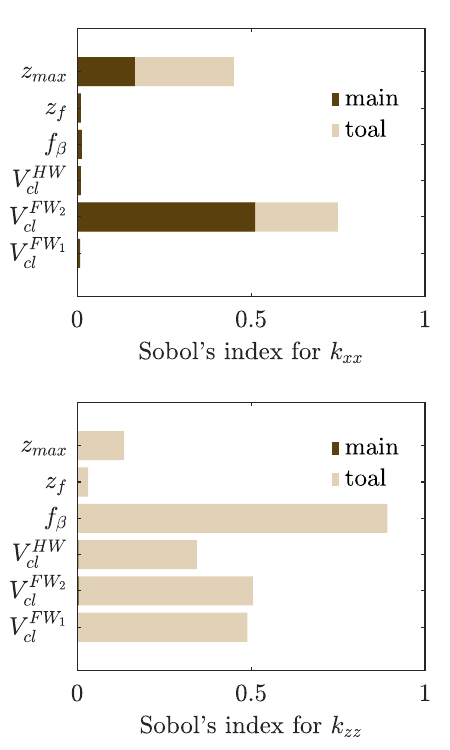}
    \caption{Sobol' indices of the 10th (left), 50th (middle), and 90th (right) percentiles for $k_{xx}$ (top row) and $k_{zz}$ (bottom row) in the SCS scenario using $N_\text{mc} = 5{,}000$ and $N_\text{sim} = 1000$. Dark brown bars represent main-effect index $S_i$, while light orange bars represent total-effect index $S_{T_i}$ calculated by Eq.~\eqref{eq:sobol}.}
    \label{fig: Sobol-SCS}
\end{figure}

\clearpage

\bmhead{Acknowledgements}
This work was funded by ExxonMobil through the ExxonMobil–MIT collaborative project ``Modeling and Mitigation of Induced Seismicity and Fault Leakage during CO2 storage."

\bibliographystyle{sn-mathphys-num}
\bibliography{main}
\end{document}